\documentclass{aastex}
\usepackage{natbib}
\usepackage{enumitem}% http://ctan.org/pkg/enumitem
\usepackage[T1]{fontenc}
\usepackage{wasysym}
\newcommand{\eg}{{\it e.g.}}
\newcommand{\ie}{{\it i.e.}}
\newcommand{\etc}{{\it etc.}}
\newcommand{\etal}{{\it et~al.}}

\newcommand{\aei}{(a,e,i)}

\newcommand{\st}{$^{st}$}

\newcommand{\km}{\,\mathrm{km}}
\newcommand{\kms}{\,\mathrm{km}/\mathrm{s}}

\newcommand{\meter}{\,\mathrm{m}}

\newcommand{\yr}{\,\mathrm{yr}}

\newcommand{\mags}{\,\mathrm{mag}}

\newcommand{\kg}{\,\mathrm{kg}}

\newcommand{\gps}{\ensuremath{g_{\rm P1}}}
\newcommand{\rps}{\ensuremath{r_{\rm P1}}}
\newcommand{\ips}{\ensuremath{i_{\rm P1}}}
\newcommand{\zps}{\ensuremath{z_{\rm P1}}}
\newcommand{\yps}{\ensuremath{y_{\rm P1}}}
\newcommand{\wps}{\ensuremath{w_{\rm P1}}}

\newcommand{\PSone}{\protect \hbox {Pan-STARRS1}}
\newcommand{\psonestar}{PS1$^*$}

\newcommand{\PSfour}{\protect \hbox {Pan-STARRS4}}

\begin{document}

%\title{Searching for small NEOs and spacecraft mission targets\\
%with the ATLAS \& \PS\ surveys. \\}

\title{The size-frequency distribution of $H>13$ NEOs\\
and ARM target candidates detected by \PSone}

\author{
Eva Schunov\'a - Lilly\altaffilmark{1} (lilly@ifa.hawaii.edu)\\
Robert Jedicke\altaffilmark{1},
Peter Vere\v{s}\altaffilmark{2},
Larry Denneau\altaffilmark{1}
Richard J. Wainscoat\altaffilmark{1}
}

\slugcomment{43 Pages, 11 Figures, 0 Tables}

\altaffiltext{1}{Institute for Astronomy University of Hawaii, 2680 Woodlawn Dr, Honolulu,
  HI, 96822, USA}
%\altaffiltext{2}{Department of Astronomy, Physics of the Earth and
  %Meteorology, Comenius University, Mlynsk\'a dolina, Bratislava, 942
  %48, Slovakia}
\altaffiltext{2}{ Jet Propulsion Laboratory, California Institute of Technology, Pasadena, CA 91109, USA}

\shorttitle{The SFD of NEOs and ARM targets detected by \PSone\.}

\shortauthors{Schunov\'a - Lilly \etal}

\begin{abstract}

We determine the
absolute magnitude (H) distribution (or size-frequency
distribution, SFD; $N(H) \propto 10^{\alpha H}$ where $\alpha$ is the slope of the distribution) for near-Earth objects (NEO) with $13<H<30$ and Asteroid Retrieval Mission (ARM) targets with $27<H<31$ that were detected by the 1\st\ telescope of the
Panoramic Survey Telescope and Rapid Response System \citep[\PSone;
  \eg][]{Kaiser2002,Kaiser2004,Hodapp2004}.  
The NEO and ARM target detection efficiencies were calculated using the \citet{Greenstreet2012} NEO orbit distribution.  The debiased
\PSone\ NEO absolute magnitude distribution is more complex than a single slope power law - it shows two transitions - at H$\sim$16 from steep to shallow slope, and in the $21<H<23$ interval from a shallow to steep slope, which is consistent with
other recent works \citep[\eg][]{Mainzer2011c,Brown2013,Harris2015}.
We fit $\alpha = 0.48\pm0.02$ for NEOs with $13<H<16$, $\alpha = 0.33\pm0.01$ for NEOs with $16<H<22$,
and $\alpha = 0.62\pm0.03$ for the smaller objects with $H>22$. There is also another change in slope from steep to shallow around H=27.
The three ARM target candidates detected by \PSone\ in one year of surveying have a corrected SFD with slope $\alpha = 0.40^{+0.33}_{-0.45}$.

We also show that the window for follow up observations of small (H$\gtrsim$22) NEOs with the NASA IRTF telescope and Arecibo and Goldstone radars are extremely short - on order of days, and procedures for fast response must be implemented in order to measure physical characteristics of small Earth-approaching objects. CFHT's MegaCam and \PSone\ have longer observing windows and are capable of following-up more NEOs due to their deeper limiting magnitudes and wider fields of view. 

\end{abstract}
%\maketitle


%\clearpage

\section{Introduction}
\label{s.Introduction}

The near-Earth objects (NEOs), asteroids or comets orbiting the Sun
with perihelion distance $q<1.3$~AU, are both
potentially threatening and beneficial --- the orbits of many NEOs
bring them close to Earth so they may eventually impact our planet,
but this also makes them the most easily accessible of all the objects
in the Solar system for  commercial (\eg\ asteroid mining) and
scientific purposes. In this paper we calculate \PSone's \citep[\eg][]{Kaiser2004,Hodapp2004} NEO and Asteroid Retrieval Mission \citep[ARM;][]{Abell2015} target detection efficiency and use it to determine an unbiased number distribution of both types of objects as a function of their absolute magnitude.

Attention has recently shifted from large NEOs with diameters $>1\km$ to smaller NEOs ($<100\meter$
diameter) as they now harbor the residual Earth impact risk and because they
are particularly convenient space mission targets.  The potential of even small NEOs to cause localized damage must not be
underestimated --- the Chelyabinsk airburst of an $\sim17\meter$
diameter object over Russia on 15 February 2013 injured about 1500
people \citep[\eg][]{Brown2013,Borovicka2013} and the Tunguska event
in 1908 flattened trees over $\sim2000\km^2$
\citep[\eg][]{Andreev1993}.

Relatively little is known about small NEOs as they are usually faint due to their
small size and thus difficult to discover unless they make a close
Earth approach. While more than 90\% of NEOs with $D>1\km$ are known,
only $\sim25$\% of those with $D>100\meter$ have been discovered and the population completeness drops down to 10$^{-5}$ for D~$\sim10\meter$ objects \citep{Harris2015,Mainzer2011a}.

 \citet{Brown2002} derived the small NEO SFD from $8.5\yr$
of satellite detections of airbursts (explosions produced by small
asteroids with $D<10\meter$ entering the atmosphere).  Their results
are consistent with the SFD extrapolated from lunar cratering records
\citep{Ivanov2006,Ito2010}.
Recently, \citet{Brown2013} suggest that a steeper slope in the SFD in this size
range might be evidence
for non-equilibrium in the population of small Earth impactors.  This would imply that the number of impactors with D~$\sim$~10-50$\meter$ may be an order of magnitude higher than estimates based on
other techniques. On the other hand, \citet{Harris2015} argue this claim should be dismissed because their results based on survey simulations agree with \citet{Brown2013} to a factor of two for this size and their SFD estimates are likely accurate within the factor of three over the whole NEO size range.

The knowledge of composition and internal structure of small NEOs is very fragmentary due to the lack of data. Smaller NEOs were thought to be monolithic rocks based on the observational data suggesting they
can rotate faster than the stability limit of a gravitationally bound
rubble pile\citep[\eg][]{Pravec2000}. However, \citet{Scheeres2010} suggest that some may still be rubble piles because cohesive Van der Waals forces
between small grains are capable of holding them together. These fast-rotating small NEOs may be the terminal state of NEO evolution driven by the YORP\footnote{The Yarkovsky-O'Keefe-Radziebskii-Paddack (YORP) effect can alter the spin rate of
irregularly shaped objects under the influence of asymmetric thermal
re-radiation} effect
\citep[\eg][]{Bottke2002b}. This theory is supported by \citet{Mommert2014a,Mommert2014b} 
who obtained infrared  measurements and high precision astrometry of two small NEAs, 2009~BD 
and 2011~MD.  The estimated bulk densities of these objects ranges from about 600 to 1800$\kg \meter^{-3}$ suggesting that both asteroids are indeed rubble-piles.

Finally, \citet{Brown2016} finds there is more than an order of magnitude spread in the strength of small NEOs impacting Earth and suggests they may be best considered as a continuum which extends from very strong monolithic objects to weakly bound sand castles.

Small NEOs are ideal objects for a detailed geological study as they can be relatively easily retrieved as samples. Particularly interesting are objects on low inclination, low eccentricity orbits with semi-major axis similar to Earth. The cratering record measurements on the Moon suggest the existence of a population of low speed projectiles \citep{Ito2010}, which might contain suitable targets for \eg\ NASA's ARM mission. 

NASA has now decided to retrieve a large 
boulder from an asteroid and deliver it to lunar orbit \citep{Abell2015} but an alternative mission design was to identify, rendezvous with, and redirect an entire small NEO of $\sim10\meter$ diameter to lunar orbit. The target population for the latter ARM mission scenario is discussed in \S\ref{ss.ARM-target-model} and referred to as `the ARM target population' because these objects remain scientifically interesting and could be candidates for future {\it in situ} resource utilization missions.  The ARM targets need to be accessible with low $\Delta v$ and must be neither too small nor too large in order to be effective resources.

A specific problem regarding the small NEO population is the short time available for follow up observations and characterization after the discovery. Due to their typically high apparent motion follow up observation is necessary within hours otherwise the orbital uncertainty grows too rapidly and the object is inevitably lost. A short observing window also makes the physical characterization of small NEOs challenging, requiring rapid response in scheduling the observation and possible cooperation between observatories \citep{Mommert2016,Thirouin2016}. \S\ref{ss.Follow-up windows} describes this issue in greater detail and concludes with our measurement of the observing windows of NEOs and ARM targets as a function of the absolute magnitude for multiple optical, IR telescopes and radar facilities.

\section{Method}
\label{s.Method}

To determine the actual NEO population's absolute magnitude distribution for $13<H<30$ and the ARM target population with $27<H<31$ we developed a synthetic data processing pipeline to measure the performance of
a simulated \PSone\ survey (denoted as \psonestar\ through the rest of the paper).  First,
 we generated a synthetic population according to the
\citet{Greenstreet2012} NEO model.  We generated independent populations of NEOs and ARM targets in 1.0 and 0.5 magnitude wide $H$ bins respectively and assigned objects within each bin an absolute
magnitude according to the \citet{Brown2002} NEO SFD.  The number of objects generated in each bin was sufficient to ensure good statistics in the synthetic discovered population.  To reduce the processing requirements NEOs and ARM targets that met a
size-dependent Minimum Orbit Intersection Distance (MOID) requirement
with Earth were injected into the \PSone\ Moving Object Processing System
\citep[MOPS;][]{Denneau2013} to simulate the survey.  The synthetic 
detections identified by MOPS were then post-processed to account for 
tracklet identification efficiency and trailing losses to mimic the 
real \PSone\ survey. The remaining synthetic detections were then 
used to calculate the \psonestar\ NEO and ARM target identification efficiency.  
These steps are described further in the following sections.

\subsection{NEO Model}
\label{ss.NEO-model}

The synthetic NEO orbit distribution was generated according to the
\citet{Greenstreet2012} model that corrects several deficiencies of
the long-standing \citet{Bottke2002a} NEO model including 1) having higher
resolution in semi-major axis ($a$), eccentricity ($e$), and
inclination ($i$), 2) using higher statistics integrations, with 3) a
finer time resolution, and 4) incorporating retrograde NEOs.  The first
point is particularly important to the detection of NEOs accessible to
human space missions because the \citet{Bottke2002a} NEO model
includes relatively few bins covering the range of ARM target
candidate orbits. 

\begin{figure}
\centering
\includegraphics[width=0.7\textwidth]{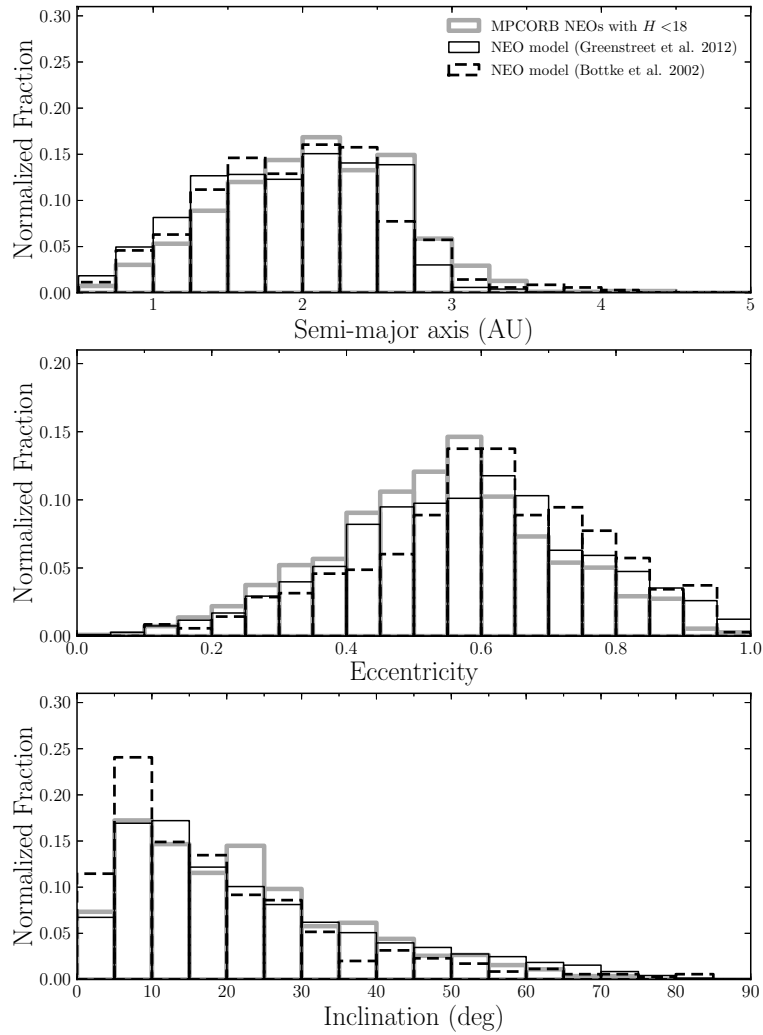}
\caption{Semi-major axis, eccentricity and inclination distributions
  of known NEOs with $H<18$ and the predicted distributions from the
  \citet{Bottke2002a} and \citet{Greenstreet2012} models.  
  }
\label{fig.NEO-model}
\end{figure}

The \citet{Greenstreet2012} NEO model uses the same weighting as the
\citet{Bottke2002a} model for the different main belt NEO sources but
their finer time and orbital element resolution results in
some important differences between the orbit distributions
(fig.~\ref{fig.NEO-model}).  A Kolmogorov-Smirnov (KS) comparison of both models with the $\aei$ distributions of known NEOs with $H<18$ (a nearly complete population) reveals that neither model formally agrees with the observed NEO distribution at the 95\% confidence level.  Despite the formal disagreement, the \citet{Bottke2002a} model's semi-major axis distribution agrees better with the observed NEOs while the \citet{Greenstreet2012} model is a better match in both eccentricity and inclination.  Furthermore, our calculation of the detection efficiency as a function of absolute magnitude is relatively insensitive to the details of the underlying `hidden' orbit distribution, and the uncertainties on the debiased SFD will be limited by the detection statistics, not the underlying model's systematic errors.

It is more difficult to compare the orbit element
distributions at smaller sizes because of observational selection
effects but \citet{Mainzer2011b} suggest that the models under-predict 
the number of objects on low-inclination orbits.
\citet{Ito2010}  also suggest the existence of a previously undetected population of slow, 
low-speed impacting objects responsible for the rayed crater distribution 
on the Moon (both works set no restrictions on the size of the objects). 
Furhtermore, current NEO models do not include the contribution of lunar ejecta. Large impacts on the Moon could have provided a population of small low $\Delta v$ NEOs akin to 1991 VG \citep{Tancredi1998}, which would be good targets for space mission (\eg\ ARM).

This work focuses on NEOs with $H>13$ ($D \la 8.5\km$)\footnote{\,unless otherwise stated
 we use a mean geometric albedo of $\rho = 0.15$ from {\tt http://sbn.psi.edu/pds/resource/albedo.html}}  
 observed by \PSone\ during calendar years 2014 and 2015 when operations were stable and the system was devoted nearly 100\% to NEO surveying. 
 We limit our study to $H < 30$ ($D \ga 
5\meter$) because 1) we know that the detection efficiency is
exceedingly small for objects in this size range (as of August 2016 there are
only 26 known NEOs with $H>30$), 2) this size overlaps the range of the
largest bolides \citep{Brown2016}, and 3) it overlaps with
our ARM target sample (see section~\S\ref{ss.ARM-target-model}).

We generated synthetic NEOs in each of 17 one magnitude-wide bins in
the $13 \le H < 30$ range and assigned each an absolute magnitude {\it within the bin} with a probability $\propto 10^{0.54\,H}$ \citep{Brown2002}. The shape of the $H$ distribution within a bin is not particularly important because we will calculate the detection
efficiency on a bin-by-bin basis, but our $H$ bins
are relatively wide so the \citet{Brown2013} SFD was employed to increase the fidelity of the efficiency calculation in the $H>25$ bins where the efficiency changes most quickly.  We generate enough synthetics
in each bin that the simulated survey (\S\ref{ss.PS1-SurveySimulations})
would detect $>100$ objects per bin (before implementing system losses)
so the uncertainty on the corrected number of \psonestar\ objects is not
dominated by the statistical uncertainty in our efficiency
determination.  This required the generation of 300 million synthetic
objects in the bins corresponding to the smallest objects with the
lowest detection efficiency.

\subsection{ARM target model}
\label{ss.ARM-target-model}

The ARM target population is dynamically restricted by five
criteria\footnote{\,\label{footnote.ARMspecs} Chodas \etal, JPL,
  (personal communication)}:
\vspace{-12pt}
\begin{enumerate}[label=\roman*)]
\itemsep-0.5em
\item 0.7~AU $<q<$ 1.05~AU
\item 0.95~AU $<Q<$ 1.45~AU
\item $2.99233<T_E<3.01$
\item $e > -1.40591 {a\over{\tt AU}} + 1.33562$
\item $e > +0.89132 {a\over{\tt AU}} - 0.93588$
\end{enumerate}
\vspace{-12pt}
where $q=a(1-e)$ and $Q=a(1+e)$ are an object's perihelion and
aphelion distances respectively, and $T_E$ is its Tisserand
parameter with respect to Earth,
\begin{equation}
T_E = {{\tt AU} \over a} + 2 \cos i \; \sqrt{{a\over{\tt AU}} (1 - e^2)}.
\vspace{-12pt}
\end{equation}
Their dynamical limits naturally restrict the objects' accessibility for a spacecraft mission as measured by the candidate's $v_{\infty}$ and the required mission $\Delta v$. We use \citet{Shoemaker1978} to estimate the $\Delta v$ required to rendezvous with the object in two maneuvers beginning from low-Earth orbit, and estimate the object's excess speed above Earth's escape speed$^{\ref{footnote.ARMspecs}}$ with $v_{\infty} \sim 29.76 \sqrt{3-T_E} \km \sec^{-1}$.  As of  August 2016 there were only 41 known objects meeting all the requirements described below.

\begin{figure}
\centering
\includegraphics[width=\textwidth]{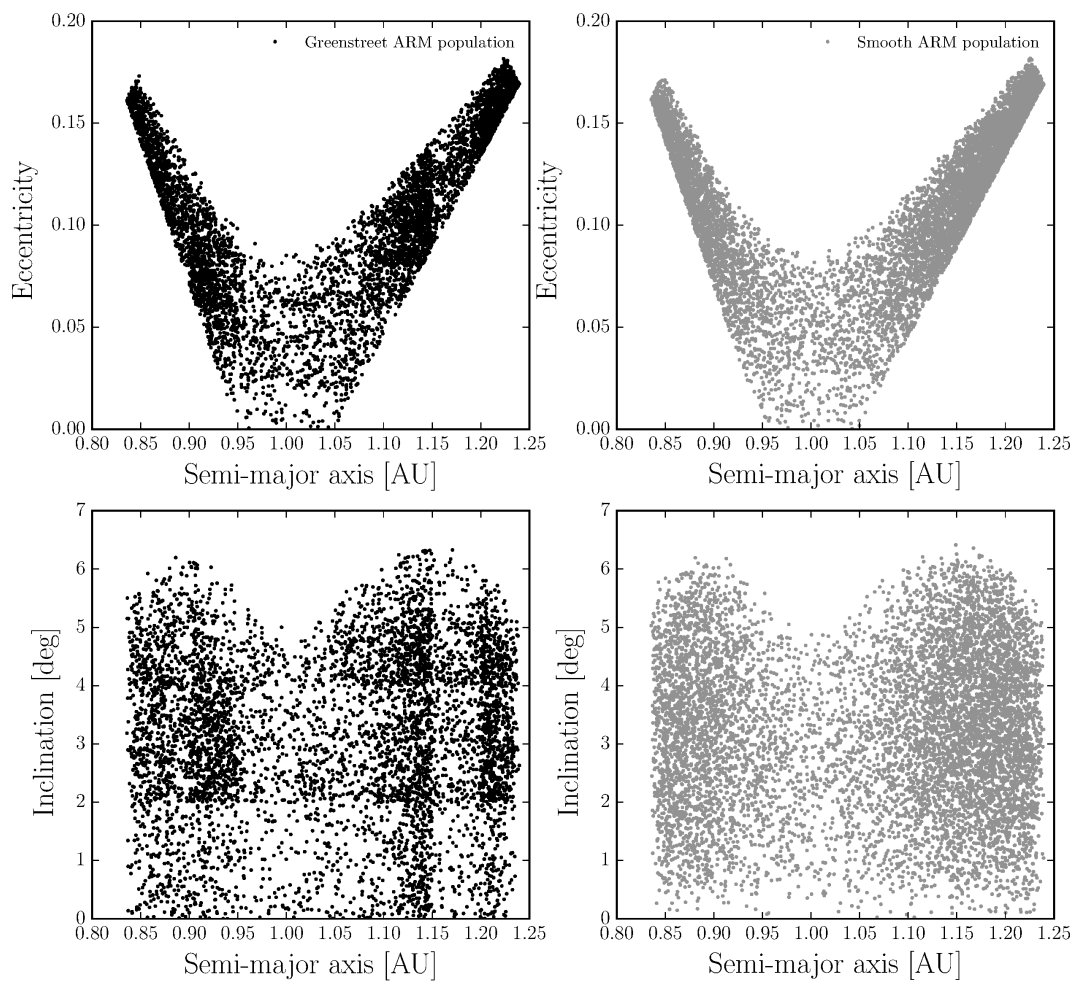}   
\caption{(top) Eccentricity vs. semi-major axis and (bottom) inclination vs. semi-major
  axis for synthetic ARM targets generated according to the raw (left) and  smoothed (right)
  \citet{Greenstreet2012} NEO model (\S\ref{ss.ARM-target-model}).  Note the
  limited range of each element and the disappearance of bin edges in the smoothed population.}
\label{fig.Greenstreet-pop1}
\end{figure}

The ARM targets' Earth-like orbits and small
sizes make them distinct from the generic NEO
population described above (\S\ref{ss.NEO-model}).  Only about 0.0030\% of NEOs in the \citet{Greenstreet2012} model pass
the ARM target selection cuts, yielding a population of objects on
extremely Earth-like orbits (fig.~\ref{fig.Greenstreet-pop1}) with
$a\sim1$~AU, small eccentricities ($e\la0.15$) and low inclinations
($i\la6\arcdeg$).  Due to the selection criteria 73\% of the ARM
targets have orbits with $\Delta v <5\km \sec^{-1}$ and all of them
have $v_{\infty} < 3.6\km \sec^{-1}$, both conditions being required
for a fuel- and cost-efficient spacecraft-asteroid rendezvous and
return mission.  For comparison only 0.2\% of the known NEO population has
$\Delta v <5\km \sec^{-1}$ and only 0.3\% of them fulfill the
condition that $v_{\infty} < 3.6\km \sec^{-1}$.

The \citet{Greenstreet2012} NEO
model has only 8 $a$-bins, 8~$e$-bins
and 3~$i$-bins spanning the range of the ARM targets's orbital elements so
that the generation of random synthetic objects according to the model created
a strong and unphysical signature of the bin edges in the resulting
$\aei$ distributions (fig.~\ref{fig.Greenstreet-pop1}). This would have an unrealistic effect on the predicted ARM 
target detection rates due to \eg\ clustering of objects along the bin edges, and artificially enhancing the number of objects with $i\sim0\arcdeg$. We 
overcame this difficulty by fitting the \citet{Greenstreet2012} NEO model to empirically selected functions in the 
192 ($=8 \times 8 \times 3$) $\aei$ bins corresponding to the ARM targets.  First, we verified that the distributions in each parameter are roughly independent and then the semi-major axis distribution was fit to a quartic function of the form
\begin{equation}
f_a(a) = -893.337 + 3563.68 \; ({a/{\tt AU}})^1 - 5141.61 \; ({a/{\tt AU}})^2 + 3204.28 \; ({a/{\tt AU}})^3 - 724.582 \; ({a/{\tt AU}})^4,
\end{equation}
the eccentricity to a gaussian of the form
\begin{equation}
f_e(e) = 43.9115 \; \exp^{-{1 \over 2}\bigl({e-0.184212 \over 0.067361}\bigr)^2},
\end{equation}
and the inclination to a line forced to be zero at $i=0\arcdeg$
\begin{equation}
f_i(i) = 13.7509 \; ({i/{\tt deg}})^1.
\end{equation}
The normalization of each of the functions is immaterial when generating random $a$, $e$ and $i$.  We then generated random sets of $\aei$ and applied the ARM target dynamical cuts to select objects from our smoothed sample.  This `smoothed' model allowed us to rapidly generate synthetic ARM targets and eliminated the bin edge effects (fig.~\ref{fig.Greenstreet-pop1}).

The ARM targets's absolute magnitudes are restricted to the range $27
\le H < 31$ corresponding to diameters $2<D< 30\meter$
depending on their albedo.  We generated objects in eight
0.5-magnitude wide bins in this size range in the same manner as for
the NEOs described above using the \citet{Brown2002} NEO SFD.  We used
0.5 magnitude wide bins because we expect the efficiency to quickly drop somewhere in this range.

\subsection{MOID selection}
\label{ss.MOIDSelection}

The fraction of all NEOs and ARM targets that can be detected with
ground-based systems decreases as the size of the object decreases
($H$ increases) because only those that approach close to Earth get
bright enough to be brighter than a survey's limiting apparent
magnitude ($V_{limit}$).  To reduce the computation time for the
survey simulations we eliminated all the generated objects with MOID
\citep[\eg][]{Gronchi2005} greater than the maximum distance at which
an object with the generated absolute magnitude can be detected by a
survey.  The maximum distance is (usually) determined by the object
having its brightest apparent magnitude ($V$) when fully illuminated
at opposition (\ie\ zero phase angle, $\alpha$).  Thus, we select
objects from our two generated populations that satisfy $
V(H,\Delta={\tt MOID},\alpha=0) < V_{limit}$ before running them
through the survey simulation.  This is equivalent to requiring that
${\tt MOID}<\Delta(H)$ where $\Delta(H)$ is the $H$-dependent maximum
distance at which the object can be detected.

Essentially all the generated NEOs with $H\la21$ are bright enough to
be detected by \PSone\ if they happen to be at their MOID while near opposition (but most objects will not satisfy these conditions)
and the fraction that pass the MOID selection criterion decreases with
the size of the NEO.  About 36\% of NEOs satisfy the MOID cut for
\PSone\ at $H\sim27$, the largest size for our synthetic ARM targets.

Roughly $2-3\times$ the fraction of ARM targets pass the MOID
criterion compared to NEOs at the same absolute magnitude because of
the ARM targets's restricted dynamical criteria.  The ARM targets's
Earth-like, low-eccentricity, low-inclination orbits with $a \sim
1$~AU are specifically designed to bring them close to Earth.

\subsection{\PSone}
\label{ss.PS1}

$\PSone$ \citep[\eg][]{Kaiser2004,Hodapp2004} was intended to be a
single-telescope prototype for a next-generation survey system
\PSfour\ \citep{Kaiser2002}. It is situated in the USA on the summit of
Haleakala, Maui, Hawaii (observatory code F51) and has been operated by the
University of Hawaii since the spring of 2010.  \PSone\ has a
$1.8\meter$ diameter primary mirror with $\sim7\deg^2$ field of
view and can survey $\sim900\deg^2$/night for asteroids (\ie\ imaged
$4\times$/night).  The \PSone\ detector system is outfitted with six
optical filters ($\gps, \rps, \ips, \zps, \yps, \wps$), where the \wps\ filter has a wide bandwidth
($\sim\gps+\rps+\ips$) optimized for asteroid detection and the \zps\ filter 
is a near-IR filter suitable for high sky-background 
 conditions, \eg\ surveying close to the Moon. 

We simulated the \PSone\ survey from the beginning of 2014 when the system was nearly 100\% devoted to NEO discovery and the bore sites were arranged mostly near the ecliptic and towards opposition to
maximize the likelihood of NEO detection.  The bore sites were selected to give a wide berth to the Moon and the galactic
plane to avoid regions with scattered moonlight and high stellar sky-plane density.  The survey pattern includes
a `sweet spot' component to enhance the detection rate of Potentially
Hazardous Objects (PHO), asteroids and comets with $H<22$ and MOID$<0.05$~AU.  The sweet spots are patches of sky roughly
 $60\arcdeg$ to $90\arcdeg$ from the Sun and within $20\arcdeg$ of
the ecliptic where the sky-plane density of PHOs is highest
\citep{Chesley2004}.  The average `transient time interval' (TTI), the time between repeated
visits to the same footprint within a night, is
$\sim19$~minutes.  

Every morning the image processing pipeline
\citep[IPP;][]{Magnier2006} produced a source list of transient
detections that included both real objects and false detections. This
list is then processed by the Moving Object Processing System
\citep[MOPS;][]{Denneau2013}.  MOPS is an integrated system that
processes data from per-exposure transient detection source lists to
identify moving objects and produces `tracklets', sets of two or more detections which correspond to the same object.

\subsection{\PSone\ survey simulation - \psonestar\ }
\label{ss.PS1-SurveySimulations}

We used MOPS to simulate \PSone\ performance and refer to the simulation as the
\psonestar\ survey.  MOPS ingests a list of bore sites and times,
the size and shape of field of view (FOV), the list of orbital elements and absolute
magnitudes of synthetic objects, and determines which of them
appear in each field and, if they do so, their apparent magnitudes and
apparent rates of motion.  MOPS was
designed to account for many additional survey factors like identification detection efficiency, camera focal-plane fill-factor, weather,
\etc, but in our work we accounted for these factors in post-MOPS
processing as described in \S\ref{ss.SystemLosses}.

\psonestar\ uses the actual bore sites 
visited by \PSone\ during the time period from 1 January 2014 through 31 May 2015. We selected only bore sites 
imaged in the \wps\ filter that is optimized for NEO discovery 
(about 80\% of all NEOs were detected in this band). 

\psonestar\ has a fixed limiting magnitude of $V_{limit}=22.5$, with 100\% detection
efficiency for $V<V_{limit}$ and 0\% for $V\ge V_{limit}$, dramatically different from the narrow and smooth drop in detection efficiency for the actual system.   To compensate for this difference we implemented an {\it ad hoc} post-process correction to the \psonestar\ $V$ distribution to ensure that it would match the observed $V$ distribution for real \PSone\ objects.  First, we fit the \PSone\ survey's asteroids's $V$-distribution to the function 
\begin{equation}
F (V) = { -0.54+ 0.04\,
V \over 1 + \exp^{(V - 21.81)/0.18} },
\end{equation}
where the numerator then provides a measure of the actual increase in number of detected objects as a function of apparent magnitude, but the denominator measures the detection efficiency near the system limiting magnitude:
\begin{equation}
\epsilon(V) = { 1 \over 1 + \exp^{(V - 21.81)/0.18}) }.
\end{equation}
We then implemented this efficiency function by randomly rejecting a fraction of detections at each limiting magnitude to ensure that our synthetic $V$ distribution matched the actual one.

\subsection{System losses}
\label{ss.SystemLosses}

The real \PSone\ survey performance is affected by \eg\ weather and system downtime, 
cosmetic imperfections in the detectors, and by so-called `trailing losses' in which fast-moving asteroids leave trailed images on the CCD rather than PSF-like detections. 
Since our \psonestar\ survey used the actual \PSone\ bore 
sites there was no need for a weather-related downtime adjustment but
we accounted for the following effects that have 
an impact on the NEO detection efficiency in post-processing:

\subsubsection{Tracklet identification efficiency}
\label{sss.TrackletEff}
Tracklet identification efficiency is a combination of several factors
including the camera's fill factor, the stellar sky-plane density,
image processing detection efficiency, \etc\   The realized \PSone\ tracklet efficiency even at bright magnitudes is about
75\% due mostly to the fill factor --- about 25\% of the
\PSone\ camera focal plane is occupied by gaps between the CCDs, gaps
on the CCDs themselves, and inactive pixels.  To mimic this loss 
we randomly deleted 25\% of synthetic detections from the \psonestar\  
simulation and required that each remaining tracklet consisted of $\ge 2$ 
detections.  (\PSone's standard operating procedures require a minimum of 3 detections for a validated tracklet but not all the detections are necessarily submitted to the Minor Planet Center because some of the detections are contaminated or otherwise adversely affected by focal plane issues.  The fraction of 2-detection tracklets actually submitted by \PSone\ is comparable to the fraction of 2-detection tracklets generated by our synthetic \psonestar\ technique.)  To assess the impact of the detection-loss process in the low-statistics, large absolute magnitude regime, we repeated the procedure 1,000$\times$ to evaluate the systematic uncertainties on the final debiased NEO SFD.

\subsubsection{Trailing losses}
\label{sss.TrailingLosses}

\begin{figure}
\centering
\includegraphics[width=0.75\textwidth]{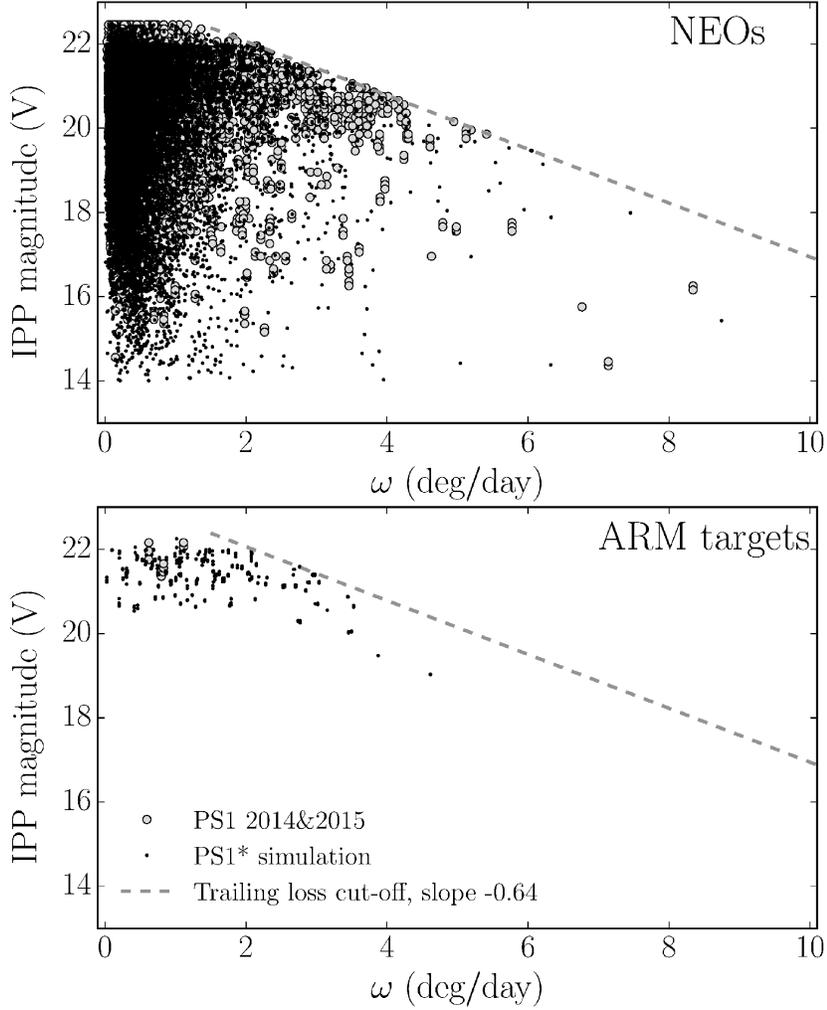}  
\caption{$V$ magnitude measured by the IPP vs. rate of motion $\omega$ of real and synthetic NEOs (top) and real and synthetic 
ARM targets (bottom) detected with \PSone\ and in our \psonestar\ simulation.   The dashed gray line represents the
empirical trailing loss limit above which asteroids are too faint to
be reliably detected.  Trailing losses begin at about $0.5\arcdeg$/day
for \PSone\ and we applied the same trailing loss limit in our
\psonestar\ simulation. 
Some \PSone\ discoveries do appear above the loss limit line due to improvements in MOPS operations but we applied the same cuts to ensure a consistent comparison between our simulation and the actual data set. 
}
\label{fig.trailing-rates-vs-Vmag}
\end{figure}

An object with high apparent rate of motion, $\omega$, can move across the field of view so fast
that it leaves a `trail' even in short exposures.  Trailing spreads the PSF causing a reduction in the per unit area apparent magnitude
and signal-to-noise, which leads to 
the failure to detect many small, fast-moving NEOs (fig.~\ref{fig.trailing-rates-vs-Vmag}). This reduction in tracklet identification efficiency 
due only to an object's apparent rate of motion is called `trailing loss' \citep[\eg][]{Tancredi1994,Veres2012}.

Trailing affects synthetic detections twice: (1) -  the MOPS \psonestar\ simulation calculates the expected integrated magnitude for synthetic detections but the \PSone\ system reports the stellar-PSF-fitted magnitude for the trail.  The resulting \psonestar\ magnitudes are brighter than they would be in reality given their apparent rate of motion; (2) - the simulation doesn't take into account the actual trailing loss and unlike the real NEOs, faint synthetic detections with high apparent rate of motion remain in the sample.

\begin{figure}
\centering
\includegraphics[width=\textwidth]{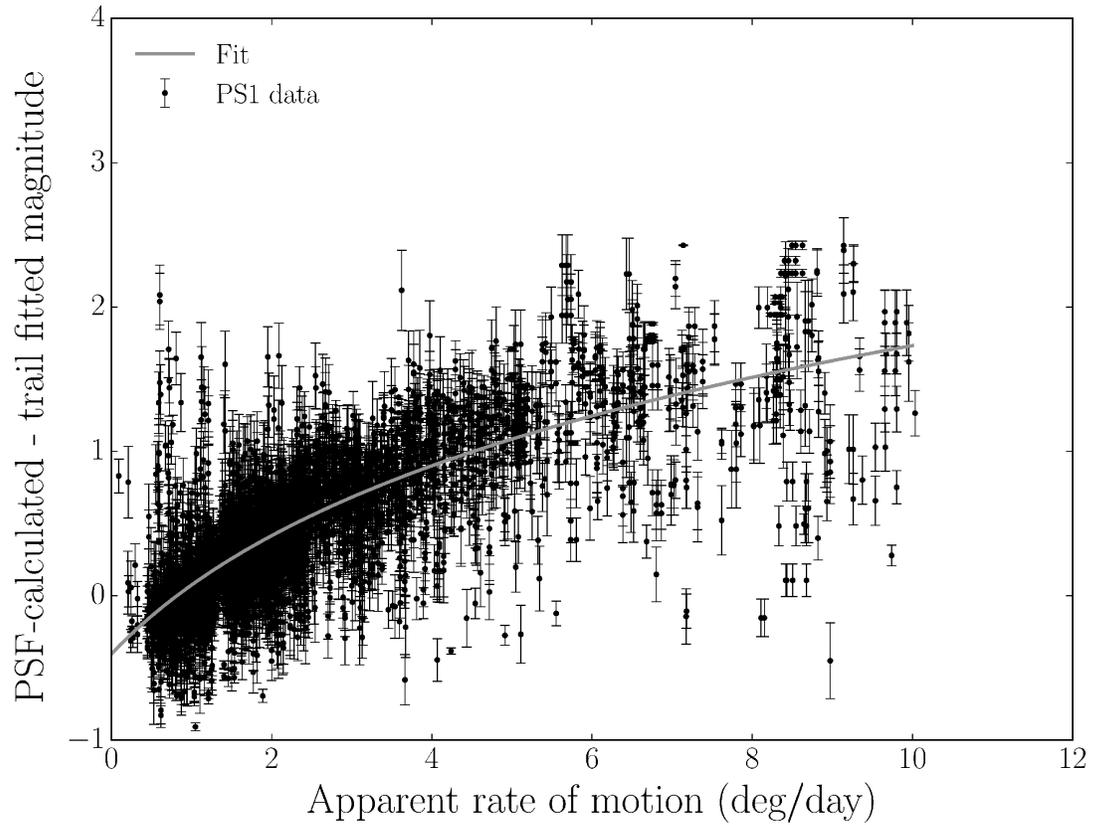}   
\caption{Difference ($\Delta$) between the \PSone\ IPP's PSF-calculated and our trail-fitted magnitudes for real detections as a function of their apparent motion ($\omega$).
}
\label{fig.trail-fitting}
\end{figure}

To account for issue (1) we `reverse trail-fit' to generate synthetic detections consistent with the real ones generated by the IPP.  \ie\ we use MOPS to generate the expected apparent magnitude, modify that value in a `reverse trail-fit', and then determine whether the trail was detected based on the modified, pseudo-realistic value.  We empirically selected a functional form for the `reverse trail-fit' based on real \PSone\ detections as illustrated in fig.~\ref{fig.trail-fitting}.  The data was fit to a logarithmic function of the form 
\begin{equation}
\Delta (\omega) = 2.79 \,\, \log_{10}(2.07 + \omega) - 1.28
\end{equation}
where $\Delta \equiv V_{IPP} - V_{trail}$, $V_{IPP}$ is the stellar-PSF-magnitude calculated by the IPP, and $V_{trail}$ is a trail-fitted magnitude. 
The trail fitting algorithm works well for fast NEOs, but fails for objects with rate of motion $\omega<$~0.7 $\arcdeg$/day, where the value of $\Delta$ becomes negative, \ie\ that the actual brightness of the object is overestimated.
To avoid this issue we assume $\Delta=$~0 for all objects slower than $\omega<$~0.7 $\arcdeg$/day during the reverse trail fitting  process.

To remove issue (2) we have simulated the trailing losses by removing the affected fast faint synthetic detections (with already corrected magnitudes) from our sample. We have empirically determined the limiting magnitude as a function of the apparent rate of motion for both \PSone\ and \psonestar\ with a simple linear function such that the effective limiting apparent magnitude is given by  
\begin{equation}
V_{limit}(\omega) < -0.64 \,\, \omega + 23.36,
\end{equation}
where $\omega>$~1.5$\arcdeg$/day.  This limiting magnitude cut was applied post facto to the \psonestar\ simulation after adjusting for the tracklet identification efficiency and `reversed' trail fit.

\begin{figure}
\centering
\includegraphics[width=0.75\textwidth]{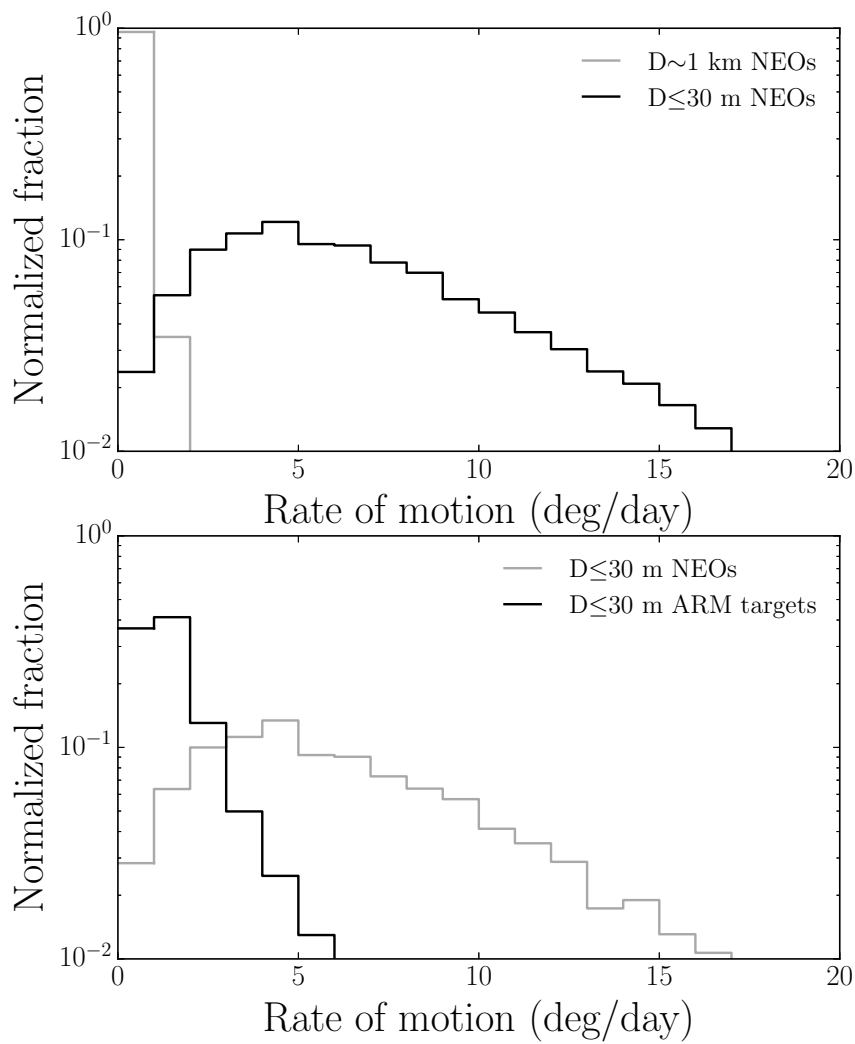}
\caption{
Rate of motion comparison between synthetic NEOs with $\sim$1 km and $\lesssim$30 m diameters ( corresponding to H$\sim$17.5 and H>27 respectively, top panel) and synthetic NEOs and ARM targets with D~$\lesssim$30 m (H>27) detected by \psonestar\ (bottom panel). The largest ARM targets
are significantly slower than NEOs at the same size. Data shown are before trailing loss implementation.}
\label{fig.NEO+ARM-ratesOfMotion}
\end{figure}

As expected, our \psonestar\ simulation shows that 
there is a dramatic difference between the apparent rates of motion of large and small NEOs 
and, interestingly also between NEOs and ARM targets at the same size 
(fig.~\ref{fig.NEO+ARM-ratesOfMotion}).  
For example, 1\,km NEOs ($H\sim17.5$) typically can be detected 
(\ie\ are brighter than the survey's $V_{limit}$) at rates of
motion of $\la2$~$\arcdeg$/day while almost all the small ones with $D\lesssim100\meter$
(H>27) are moving faster than this
rate at the time of the discovery.   This is a observational selection effect because the surveys are able to detect smaller NEOs only when they make a close 
approach to Earth which results in their high apparent rates of motion.

The ARM targets' apparent rate of motion is considerably less than the
NEOs' apparent rate of motion at the same size because the former's
Earth-like orbits ensure that they spend more time in Earth's
vicinity (and thus are slower) than generic NEOs.  For example, at
$H\sim27$ the average ARM target's apparent rate of motion when detected by \psonestar\ is
1.5~$\arcdeg$/day, while the average rate of
motion for NEOs of the same absolute magnitude is 6.7 $\arcdeg$/day. 
This suggests that objects on the ARM target-like orbits are more likely 
to be detected by surveys than `regular' NEOs, which was also noted by \citet{Harris2015}.
More importantly they are also easier targets for follow-up astrometric observations that can reduce their orbit fit residuals and ensure they won't get lost.

\subsubsection{NEO identification}
\label{sss.NEOidentification}

Contemporary asteroid surveys rely on the MPC NEO `digest' score that
is, essentially, the probability that a real tracklet is not a main-belt object (\ie\ it may be a NEO) if it
can not be associated with a known asteroid.  MPC requires that surveys
submit and follow-up NEO candidates only with digest score $\geq$~65. Objects with 
digest score lower than this threshold get submitted as `incidental astrometry'.
MPC assigns them a lower priority for processing and eventually attempts
 to find links with previous observations of known objects. If the linking routine fails, 
 tracklets are moved to the isolated tracklet file (ITF), essentially a depository of unlinked and lost detections. 
Thus, some NEOs `hide in
plain sight' by virtue of being detected but having otherwise mundane
locations and rates of motion \citep[\eg][]{Jedicke1996}.

\begin{figure}
\centering
\includegraphics[width=\textwidth]{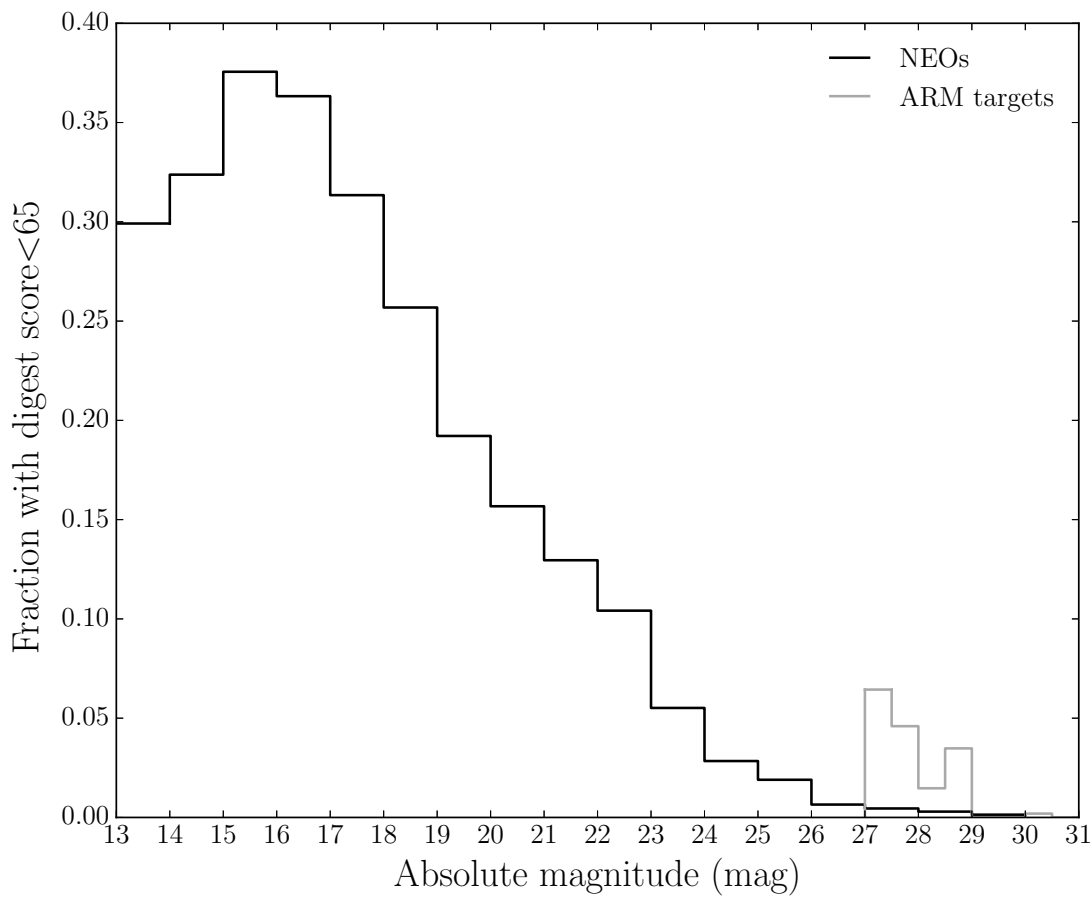}   
\caption{Fraction of NEOs and ARM targets with digest score $<$65 vs. absolute magnitude H. The fraction of NEOs detected with scores $<$65 which is the threshold of reporting them to MPC as NEOs is highest among the big bright objects that are detectable from greater distances and at lower apparent rate of motions than smaller objects. Almost 95\% of ARM targets have digest score higher then the MPC limit and are reported as NEOs.}
\label{fig.digest}
\end{figure}

Fig.~\ref{fig.digest} shows fraction of synthetic detections with digest score $<65$ per  absolute magnitude bin.
More than 30\% of the largest and almost 20\% of $\sim$1~km objects wouldn't be submitted as NEO candidates. 
However, this is not an issue for larger 
NEOs because most are already known \citep[\eg][]{Mainzer2011a, Harris2015}.
We can also expect that most of the medium-sized NEOs hiding in plain sight will be re-detected with much higher digest score over the course of a multi-year survey.
The low scores are critical for smaller objects, where a rapid follow-up is necessary otherwise the orbital uncertainly grows rapidly and they will get lost. 
Fortunately the majority of small NEOs
and ARM targets are moving so fast when detected, that
their digest score will almost always be $\sim 100$ (fig.~\ref{fig.NEO+ARM-ratesOfMotion}).
Indeed, less than 5\% of synthetic NEOs with H>22 and roughly 7\% of the largest ARM targets have scores this low and will be lost.
The fraction is slightly larger for ARM targets compared to NEOs at the same size primarily due to their slower apparent rate of motion.

\section{Results \& discussion}
\label{s.ResultsDiscussion}

\subsection{NEO and ARM target detection efficiency}
\label{ss.DetectionEfficiency}

\begin{figure}
\centering
\includegraphics[width=\textwidth]{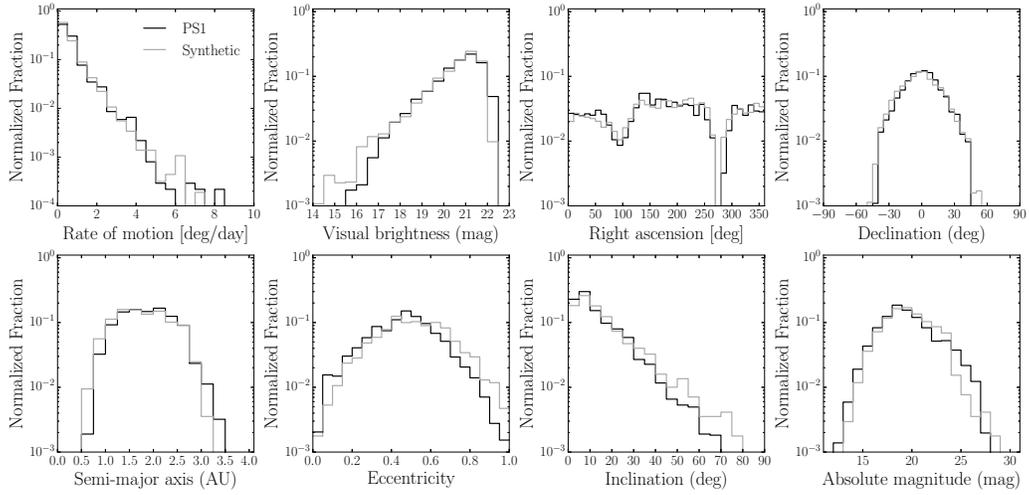}
\caption{Comparison of rate of motion, visual magnitude, RA, DEC, a, e, i and absolute magnitude of synthetic NEOs from the \psonestar\ simulation and the real \PSone\ data.}
\label{fig.NEO-PS1-vs-sim}
\end{figure}

The survey's NEO detection efficiency, $\epsilon(H)$, is the fraction of objects
identified in the simulation as a function of absolute magnitude $H$.   We combine discoveries of new objects and detections of known objects into `identifications' because we make no distinction between known and new NEOs in our simulation.  Most of the large objects will be detections, most of the small objects will be discoveries.  Fig.~\ref{fig.NEO-PS1-vs-sim}  shows our simulation is a good representation of reality and the detected synthetic NEOs match the real \PSone\ detections well in terms of rate of motion, visual magnitude, sky-plane coordinates at discovery and in (a, e, i, H).
The realistic simulated survey detection efficiency can then be used as proxy for the actual survey detection efficiency.

\begin{figure}
\centering
\includegraphics[width=\textwidth]{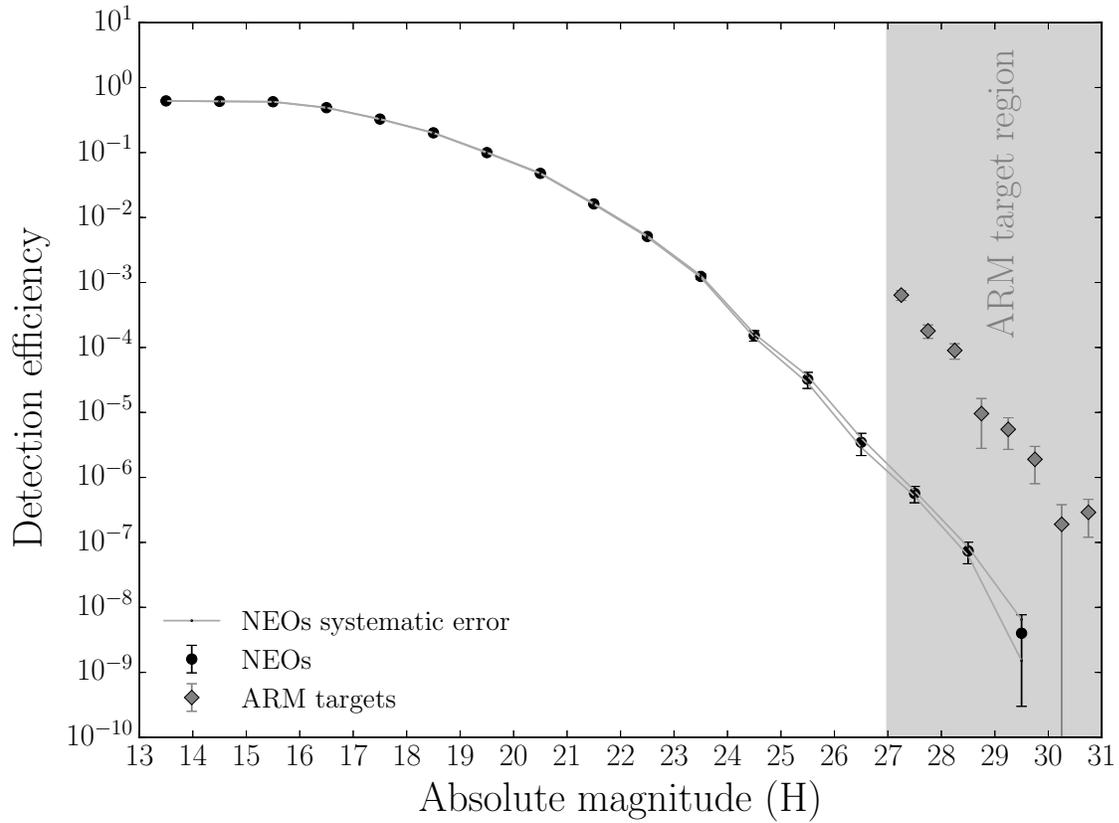}
\caption{Absolute detection efficiency for NEO and ARM targets as a function
  of absolute magnitude for the \psonestar\ simulation, \ie\ the fraction of each population that would be detected during the survey duration (January 1 2014 - May 30 2015).}
\label{fig.Annual-efficiency}
\end{figure}

We find that the survey's NEO detection efficiency decreases by about 6 orders of
magnitude as the cross-sectional area of the objects decreases by
$\sim10^4\times$ from $H=18$ to $H=28$
(fig.~\ref{fig.Annual-efficiency}).  The NEO detection
efficiency in the ARM
target size range of $27<H \lesssim31$ decreases from $\sim10^{-6}$ to $\sim10^{-9}$ because these objects are very small
and need to be close to Earth to be bright enough to be detected, but
are then moving so fast that trailing losses are high
(fig.~\ref{fig.NEO+ARM-ratesOfMotion}).

Interestingly, the ARM target detection efficiency (fig.~\ref{fig.Annual-efficiency})
is $\sim1000\times$ higher than the NEO detection efficiency {\it at
  the same absolute magnitude} because of the differences in their
orbit element distributions --- by design, the ARM targets approach
close to Earth and spend a relatively long time in Earth's environs so
they have a higher probability of being detected.  Their
slower apparent rate of motion relative to the NEOs accounts for their
higher detection efficiency.  Even so, our results suggest that only
$\sim$0.06\% of the largest ARM targets with $H\sim27$ are detected
annually by \psonestar.  Their detection efficiency drops
$\sim10^4\times$ from $H\sim27$ to $H\sim31$ --- faster than the
number of objects increases as we will show below. 

Our conclusion supports the findings of \citet{Harris2015}, even though their definition of 
ARM-target like bodies is less strict than ours (D$\sim10$~m, MOID$\lesssim0.03$~AU and $v_{\infty}$<2.5~$\kms$). 
Their results suggest the current surveys are 1000 times more efficient in detecting these slow moving objects than the 
average speed bodies and estimate their total population to 2000.

\subsection{Near-Earth object size-frequency distribution}
\label{ss.NEO-SFD}

By virtue of the similarity of our simulated \psonestar\ survey to the real \PSone\ the measured detection efficiency
(fig.~\ref{fig.Annual-efficiency}) allows us to debias the actual
\PSone\ NEO detections and derive the real NEO SFD. In each H-bin the total number of NEOs (N) is simply the number of NEOs detected by \PSone\ ($n_{PS1}$) divided by the efficiency of \psonestar\ ($\epsilon$) measured for the given bin:   $N = n_{PS1} /\epsilon$. 

\begin{figure}
\centering
\includegraphics[width=\textwidth]{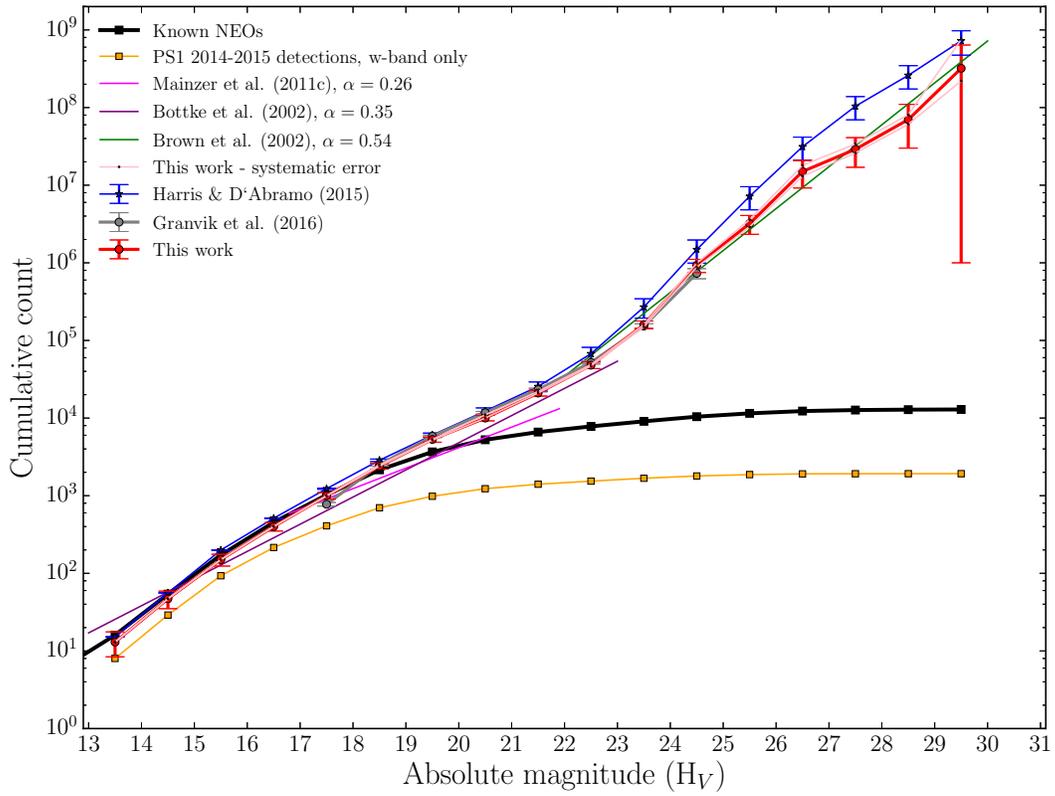}
\caption{Our derived NEO SFD from \PSone\ data (\S\ref{ss.NEO-SFD}) in
  comparison to other contemporary models and known NEOs.}
\label{fig.SFD-comparison}
\end{figure}

Fig.~\ref{fig.SFD-comparison} shows the debiased NEO SFD in comparison with other contemporary models. Our SFD is in excellent agreement with the \citep{Granvik2016} NEO model that extends the work of \citet{Bottke2002a}. All but one of our values in H-bins are within 1-$\sigma$ -- interestingly enough the largest disagreement is for H$\sim$17.5, where our prediction lies directly between the \citet{Granvik2016}  and  \citet{Harris2015} SFDs.   Our distribution also matches \citet{Harris2015} within 1-$\sigma$ up to H$=$27 but our values are systematically below their SFD beyond $H\sim22$ while we continue to be in excellent agreement with \citep{Granvik2016} to $H=25$ and \citep{Brown2002} all the way to $H\sim30$.

Our derived SFD exhibits smooth variations in the slope as a function
of absolute magnitude akin to the \citet{Harris2015} model.  To compare
the slope of the distributions we fit our cumulative SFD to a function
of the form $N(H_0) = N_{H_0} \, 10^{\alpha \, (H - H_0)}$ where
$N(H_0)$ is the cumulative number of objects with $H<H_0$,
$\alpha$ is the slope of the distribution, and $H_0$ is the absolute
magnitude at which there are $N_0$ objects with $H<H_0$.  We find $N_{13}=8.6\pm1.2$ and $\alpha = 0.48\pm0.02$ for NEOs with $13<H<16$. The slope becomes shallower for NEOs with $16<H<22$ where 
$N_{16}=353\pm34$ and $\alpha = 0.33\pm0.01$. 
Around H$=$22 the slope becomes steep again with $N_{22}=44,400\pm4,100$ and $\alpha = 0.62\pm0.03$ for NEOs with $22<H<27$ which agrees to within 1-$\sigma$ with the \citet{Brown2002,Brown2013} SFD 
slope measured for large bolides from infrasound detections.
The selection of boundaries in any non-continuous data is difficult, especially for power-law distributions \citep{Newman2005,Clauset2009}.
We empirically identified the $H$-boundaries as the values where the uncertainties on $\alpha$ and $N_{H_0}$ exceeded 5\% and 15\% respectively if they were included in the SFD. Note that the uncertainties on our SFD represented by error bars on fig.~\ref{fig.SFD-comparison} are statistical only. The systematic uncertainty introduced by detection losses due to the camera's fill factor is depicted by pink lines.

 \citet{Harris2015}  speculate that the dip around $19<H<25$ may correspond to a transition in the internal structure of asteroids from rubble piles to monolithic bodies, with its minimum around $D\sim$~100 m representing the weakest objects mostly prone to disruption. The same suggestion was made earlier by e.g. \citet{Durda1998,Bottke2005,OBrien2005}. \citet{Mainzer2011c} have already shown that there is no observed shift of mean albedo with the diameter of the object (where the transition in the SFD would represent the largest gradient of albedo change), thus the transition must have an actual structural or dynamical cause. Perhaps this transition in SFD points to the size regime where the YORP effects starts to be dominant and increases the spin rate more efficiently. This might lead to a gradual mass loss eventually leading towards more small, fast rotating NEOs \citep[\eg][]{Bottke2002b}.

Similarly to \citet{Harris2015} there is an evidence for yet another change in slope for smaller NEOs with $27<H<30$
where our fit yields a steeper slope of $\alpha =0.45\pm0.09$ with $N_{27}=16.61\pm4.04 \times 10^6$.   Our SFD is weakly constrained for $H\ga27$ because of the small number of \PSone\ detections in this size range and the larger uncertainty on
our derived efficiencies. 

We used the \citet{Harris2015} SFD to calculate the NEO detection rates expected over the duration of the survey simply because it covers the absolute magnitude range of our NEO model. Tab. 1 shows the expected rate matches the real data well up to H$\sim$22 where it starts to overestimate the number of NEOs compared to real detections by \PSone\ . This is due to the fact that values of \citet{Harris2015} SFD are systematically above ours for H$\gtrsim$22. For example \PSone\ detected  12 NEOs with $H>27$ during the analyzed time period, while we predict the detection rate of 56 NEOs if we use the SFD of \citet{Harris2015}. On the other hand, \PSone\ should detect 10 NEOs with $H>27$ when \citet{Brown2002,Brown2013} SFD is used, thus it seems to be a better match for the upper part of NEO SFD than \citet{Harris2015}.

\subsection{\PSone\ ARM target discoveries}
\label{ss.PS1-ARM-targets}

\ \PSone\ identified 3 ARM target candidates between 1 January 2014 and 31 May 2015  
while our \psonestar\ simulation predicted discovery
rates of $0.45\pm0.13$/year with the SFD calculated in this work for $27<H<30$.  Our prediction lies directly between detection rates based on the \citet{Brown2002} and \citet{Harris2015} SFDs which give $0.21\pm0.02$ and $1.1\pm0.3$ ARM targets discovered per year respectively  (tab. 2).  The detection rate of $<2$/year seems to be robust
since PS1 has detected a total of 9 ARM targets since operations began in early 2010.  A simple extrapolation of the \citet{Bottke2002a} NEO SFD to the ARM target absolute magnitude range would predict $120\times$ more than were actually identified, but this extrapolation is unwarranted given
the model's applicable range of $H<22$.

We used a maximum-likelihood (ML) fit to determine the slope ($\alpha$) of the
SFD for the 3 \PSone\ ARM targets to avoid the numerical issues of binning them
in $0.5\mags$ $H$-bins and thereby losing 
resolution on the calculated absolute magnitudes.  To do so we first fit
the ARM target detection efficiency ($\epsilon_{ARM}$; fig~\ref{fig.Annual-efficiency}) of our
\psonestar\ simulation and found
$\log_{10}\epsilon_{ARM} = (-2.96\pm0.17) - (1.04\pm0.08) \times (H-27)$ for $27<H<31$.  The
ML fit yielded an ARM target SFD $\propto 10^{\alpha H}$ with $\alpha
= 0.40^{+0.33}_{-0.45}$, suggestive of a shallow and non-equilibrium
SFD but consistent with almost anything.

The ARM targets as a dynamically restricted subset of NEOs with low $\Delta v$ may be the proposed `undetected' population of low speed lunar projectiles noted by \citet{Ito2010} from lunar cratering records. Alas, several works \citep[][\eg]{Gladman1995,Gaskin1998} suggest that some fraction of lunar ejecta will return to the Earth-Moon system within several thousand to million years. 
Some ARM targets might also be remnants from NEOs tidally disrupted during close encounters with Earth which tend to return to Earth's environs and eventually re-impact Earth within $\sim10^4$ years \citep{Schunova2014}. Finally, a small percentage are man made such as Apollo rocket boosters \citep{Chodas2013}.

\subsection{Follow-up observation windows}
\label{ss.Follow-up windows}

One issue with current \PSone\ discoveries is the lack of astrometric follow-up observations for fast and/or faint objects.  While \PSone\ can self follow-up it is better suited to rapidly surveying the sky with its wide-field camera than targeting specific objects.   This leads inevitably to the loss of many potentially interesting NEOs because there is usually only limited time available for followup observations aimed at securing the object's orbit and obtaining measurements suitable for physical characterization \citep{Mommert2016,Thirouin2016}.

Here we used synthetic NEOs and ARM targets detected by our \psonestar\ survey to assess the time windows suitable for follow up observations with the CFHT \citep{Tholen2001}, \PSone\  and NASA IRTF telescopes and also 
for the Arecibo and Goldstone radars \citep{Benner2015}.  

For each detected synthetic \psonestar\ object we calculated an ephemeris 
for 100~days after the `discovery' epoch to determine how long it would remain in the followup site's observing window in terms of RA, declination, maximum altitude during the night/day and $V_{lim}$, or how long it will be within the SNR range for the radar facilities. We assumed the minimum required radar SNR=10 for a detection\footnote{P.Chodas (NASA JPL) - personal communication}.

\begin{figure}
\centering
\includegraphics[width=0.8\textwidth]{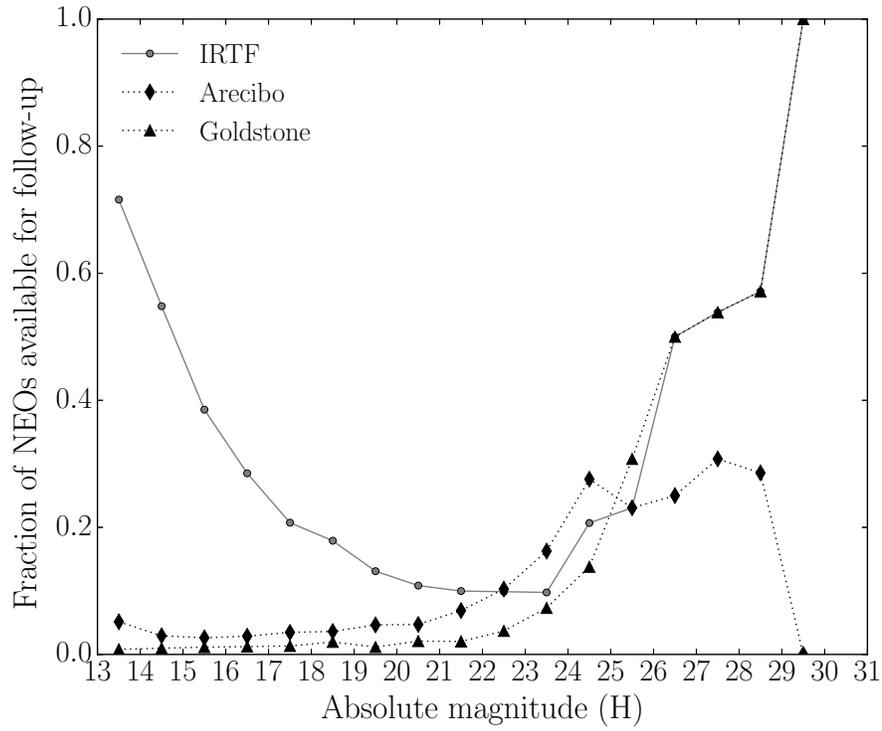}
\caption{Fraction of NEOs accessible for follow-up observations with the IRTF telescope and the Arecibo and Goldstone radar facilities for 100 days after the discovery with \psonestar\ . Virtually all NEOs are accessible to \PSone\ and CFHT after their discovery.}
\label{fig.visible-fraction}
\end{figure}

While virtually all ($>$99\%) synthetic NEOs detected by \psonestar\ were accessible for follow-up by \PSone\ and CFHT due to their fainter limiting magnitudes, many never reached sufficient brightness or SNR necessary for follow-up with IRTF,  Arecibo, or Goldstone (fig.~\ref{fig.visible-fraction}). The fraction of objects accessible for followup decreases with increasing $H$ for IRTF to $H\sim22$ while it remains flat for both radar systems in the same range. The opportunity for followup increases for all three facilities for $H\ga22$ due to observation selection effects --- small objects in this absolute magnitude range need to be close enough to be detected by \PSone and, by the virtue of their proximity, some of them become brighter than IRTF's limiting magnitude and also reach the limiting SNR of both radars. 

\begin{figure}
\centering
\includegraphics[width=0.8\textwidth]{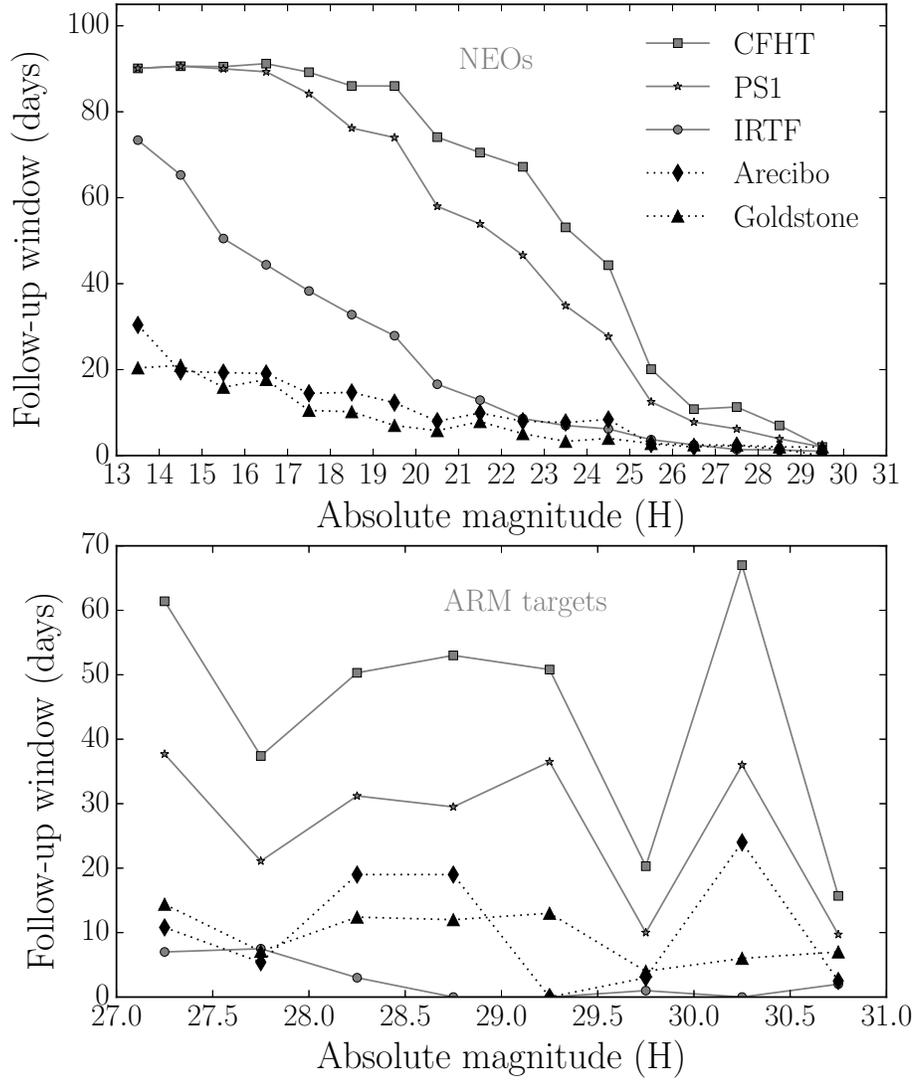}
\caption{Average follow-up window duration for NEOs and ARM targets by the NASA IRTF telescope with SPEX, \PSone\, CFHT and Arecibo and Goldstone radars after discovery by \PSone. 
}
\label{fig.obs-window}
\end{figure}

As expected IRTF/SPEX \citep{Rayner2003} has the shortest average observation window due to its lower $V_{lim}=$18.5 (fig.~\ref{fig.obs-window}). 
It is necessary to act quickly if one wants obtain a physical characterization of small NEAs and ARM targets with this instrument because these objects will be out of its reach within several days. Follow up windows get 
 progressively longer for \PSone\ and CFHT due to their deeper $V_{lim}$. \PSone\ is able to detect objects up to V=22.5 in \wps\ -band during 45~s exposure, while CFHT's MegaCam \citep{Boulade2003} is able to reach 24.0 mag in r-band during a 60 s exposure. The average windows for small NEOs are only $\sim$20 days duration for these instruments but this might be enough to obtain astrometric and photometric measurements necessary for orbit determination and physical characterization. Both CFHT and \PSone\ are equipped with cameras with large FOV ($1\deg^2$and $\sim7\deg^2$ respectively), which makes them ideal for `chasing' fast close objects. 
 
ARM targets are very difficult objects to follow-up - out of the complete set ($27<H<31$) of the synthetic ARM targets detected by \psonestar\ , only 24\% were bright enough to reach the V$_{lim}$ of IRTF and those were observable only for an average of 2.4 days. \PSone\ and CFHT would be more successful since all synthetic ARM targets were available for follow-up with these facilities and the average observing windows were 26.5 and 44.5 days respectively. Figure~\ref{fig.obs-window} shows that ARM targets have in average 2-3 times longer windows as NEOs in the same H-range no matter what facility is used which is again accountable to their much smaller apparent motion compared to that of NEOs in the same size. 
 
 The average radar windows for both both populations are typically shorter compared to the optical telescopes, which can be also accounted to the different geographic position of these facilities. In average 48.4\% and 41.0\% of ARM targets were available for follow-up with Arecibo and Goldstone radars respectively with average observing window of 10.5 and 9.5 days.

\section{Conclusions}
\label{s.Conclusions}

Our \psonestar\ simulation is a good proxy of the actual \PSone\ survey
based on the fact that it reproduces very well the (a,e,i,H) - distribution and visual magnitudes and apparent rate of motion of the real NEOs detected by \PSone\ . 

We used our normalized \psonestar\ annual detection efficiency to
debias the \PSone\ system's identified NEOs as a function of absolute
magnitude.  The resulting $H$ distribution is an excellent match to the new \citet{Granvik2016} NEO model and its shape is akin to the
\citet{Harris2015} distribution for $H<22$  where it exhibits multiple transitions between shallow and steep slopes.

Our best fit to the cumulative SFD yields $\alpha = 0.48\pm0.02$ for NEOs with $13<H<16$,  $\alpha=0.33\pm0.01$ 
for NEOs with $16<H<22$ and finally a steep slope of $\alpha=0.62\pm0.03$
for the small NEO population with $22<H<27$ which agrees within 1-$\sigma$ with the \citet{Brown2002,Brown2013} SFD 
slope measured for large bolides from infrasound detections.  Even smaller NEOs with $H>$~27
exhibit a possible turn to a shallower slope where our fit yields $\alpha =0.45\pm0.09$.

Our maximum-likelihood fit to the \PSone\ ARM target SFD yielded $\alpha =
0.40^{+0.33}_{-0.45}$, consistent with the entire range of slopes for
the known NEO population.

We show that the windows for follow up observations of small (H$\gtrsim$22) NEOs and ARM targets with NASA IRTF telescope and Arecibo and Goldstone radars are extremely short - in order of several days and the CFHT's MegaCam and \PSone\ would be more successful in obtaining astrometric and/or photometric observations of objects across the whole inspected H-range. 
Thus we recommend that procedures for fast response should be implemented in order to measure physical characteristics, such as rotation rate, shape, and spectra of small Earth-approaching objects.

\clearpage
\acknowledgements
\section*{Acknowledgments}

We thank Paul Chodas, Donald Yeomans and Steve Chesley from NASA's NEO
office at the Jet Propulsion Laboratory for defining the ARM target
population and helpful discussions about the small NEO population; Scott Stuart, MIT Lincoln Laboratory, for his insights on measuring 
the ARM targets' size-frequency distribution; the reviewers for many helpful comments and extensive discussion; and Robert Weryk from the IfA Unversity Hawaii for his thorough revision of the manuscript.
Eva Lilly thanks her newborn daughter Elizabeth for being a good baby so she could finish the paper in peace.

The Pan-STARRS1 Surveys (PS1) have been made possible through contributions of
the Institute for Astronomy, the University of Hawaii, the Pan-STARRS
Project Office,
the Max-Planck Society and its participating institutes, the Max
Planck Institute for
Astronomy, Heidelberg and the Max Planck Institute for
Extraterrestrial Physics, Garching,
The Johns Hopkins University, Durham University, the University of
Edinburgh, Queen?s
University Belfast, the Harvard-Smithsonian Center for Astrophysics,
the Las Cumbres
Observatory Global Telescope Network Incorporated, the National
Central University
of Taiwan, the Space Telescope Science Institute, the National
Aeronautics and Space
Administration under Grant Nos. NNX08AR22G, NNX12AR65G, and NNX14AM74G
issued through the Planetary Science Division of the NASA Science
Mission Directorate, the
National Science Foundation under Grant No. AST-1238877, the
University of Maryland,
and Eotvos Lorand University (ELTE) and the Los Alamos National Laboratory.

\bibliographystyle{agsm}
\bibliography{references3}

% TA B L E S -------------------------------------------------------

\clearpage
\begin{deluxetable}{cccccc}
\tablewidth{440pt} 
\tabletypesize{\footnotesize}
\tablecolumns{6}
\label{table.NEO-detection-statistics}
\tablecaption{NEO detection efficiency ($\epsilon$) and predicted number of detections for the simulated \psonestar\ survey assuming the SFD from \citet{Harris2015}, and the actual number of detections for the \PSone\ survey for the time period 2014 January 1 to 2015 May 31.}

\tablehead{
\colhead{H$_V$ range (mag)} & 
\colhead{$\epsilon \pm \Delta \epsilon$(\psonestar\ )} &
\colhead{N $\pm \Delta$N(\psonestar)} & 
\colhead{N(\PSone)} 
}
\startdata

13.0	-	14.0	&	6.23	$\pm$	0.25	$\times$	10$^{-1}$	&	6.3	$\pm$	0.3	&	8	\\
14.0	-	15.0	&	6.13	$\pm$	0.25	$\times$	10$^{-1}$	&	24.9	$\pm$	1.0	&	21	\\
15.0	-	16.0	&	6.10	$\pm$	0.25	$\times$	10$^{-1}$	&	87.2	$\pm$	3.5	&	64	\\
16.0	-	17.0	&	4.88	$\pm$	0.16	$\times$	10$^{-1}$	&	152.6       $\pm$	5.1	&	122	\\
17.0	-	18.0	&	3.26	$\pm$	0.10	$\times$	10$^{-1}$	&	233.3	$\pm$	9.4	&	194	\\
18.0	-	19.0	&	2.03	$\pm$	0.01	$\times$	10$^{-1}$	&	330.7	$\pm$	21.5	&	290	\\
19.0	-	20.0	&	1.01	$\pm$	0.00	$\times$	10$^{-1}$	&	318.6	$\pm$	30.6	&	284	\\
20.0	-	21.0	&	4.77	$\pm$	0.20	$\times$	10$^{-2}$	&	297.4	$\pm$	41.2	&	248	\\
21.0	-	22.0	&	1.63	$\pm$	0.08	$\times$	10$^{-2}$	&	215.7	$\pm$	41.6	&	174	\\
22.0	-	23.0	&	5.06	$\pm$	0.32	$\times$	10$^{-3}$	&	212.6	$\pm$	54.1	&	135	\\
23.0	-	24.0	&	1.28	$\pm$	0.12	$\times$	10$^{-3}$	&	257.9	$\pm$	81.8	&	137	\\
24.0	-	25.0	&	1.51	$\pm$	0.28	$\times$	10$^{-4}$	&	182.9	$\pm$	70.9	&	118	\\
25.0	-	26.0	&	3.25	$\pm$	0.90	$\times$	10$^{-5}$	&	185.6	$\pm$	80.3	&	74	\\
26.0	-	27.0	&	2.00	$\pm$	1.00	$\times$	10$^{-6}$	&	48.3	$\pm$	28.9	&	42	\\
27.0	-	28.0	&	5.97	$\pm$	1.65	$\times$	10$^{-7}$	&	43.4	$\pm$	18.8	&	8	\\
28.0	-	29.0	&	7.00	$\pm$	2.65	$\times$	10$^{-8}$	&	10.9	$\pm$	5.5	&	3	\\
29.0	-	30.0	&	3.34	$\pm$	3.30	$\times$	10$^{-9}$	&	1.5	$\pm$	0.0	&	1	\\
\enddata

\end{deluxetable}

\clearpage
\begin{deluxetable}{cccccc}
\tablewidth{440pt} 
\tabletypesize{\footnotesize}
\tablecolumns{6}
\label{tab.ARM-detection-statistics}
\tablecaption{Detection efficiency ($\epsilon$) and predicted number of ARM target detections for the simulated \psonestar\ survey assuming the SFD from \citet{Harris2015}, and the actual number of detections for the \PSone\ survey for the time period 2014 January 1 to 2015 May 31. \citet{Harris2015} SFD does not contain data for $30.5<H<31.0$.}

\tablehead{
\colhead{H$_V$ range (mag)} & 
\colhead{$\epsilon \pm \Delta \epsilon$(\psonestar\ )} &
\colhead{N $\pm \Delta$N(\psonestar)} & 
\colhead{N(\PSone)} 
}
\startdata
 & $\times$ 10$^{-10}$  & $\times$ 10$^{-2}$  &  \\
\tableline
27.0	-	27.5	&	191.4 $\pm$	33.3	&	52.3	$\pm$	25.4	&	1	\\
27.5	-	28.0	&	55.1	$\pm$	12.6	&	25.0	$\pm$	12.8	&	2	\\
28.0	-	28.5	&	27.1	$\pm$	7.2	&	20.1	$\pm$	10.7	&	0	\\
28.5	-	29.0	&	2.9	$\pm$	2.1	&	2.4	$\pm$	2.0	&	0	\\
29.0	-	29.5	&	1.7	$\pm$	0.8	&	3.4	$\pm$	2.4	&	0	\\
29.5	-	30.0	&	0.6	$\pm$	0.3	&	1.5	$\pm$	1.2	&	0	\\
30.0	-	30.5	&	0.1	$\pm$	0.1	&	0.4	$\pm$	0.4	&	0	\\
30.5	-	31.0	&	0.1	$\pm$	0.1	&	-	$\pm$	-	&	0	\\
\enddata
\end{deluxetable}

% F I G U R E S -------------------------------------------------------

\clearpage

\clearpage

\clearpage

\clearpage

\clearpage

\clearpage

\clearpage

\clearpage

\clearpage

\clearpage

\clearpage

\end{document}